\RequirePackage{ifpdf}
\documentclass[11pt,a4paper]{article}
\pdfoutput=1
\usepackage{jcappub}
\usepackage{amsmath,amssymb,epsf}
\usepackage{graphicx,color}

\def\bel#1{\begin{equation} \label{#1}}

\def\st{\phi_*}

\def\vp{\varphi}
\def\mpl{M_{\rm pl}}

\def\mc{\mathcal}
\def\be{\begin{equation}}
\def\ee{\end{equation}}
\def\bea{\begin{eqnarray}}
\def\eea{\end{eqnarray}}

\def\ltap{\ \raise.3ex\hbox{$<$\kern-.75em\lower1ex\hbox{$\sim$}}\ }
\def\gtap{\ \raise.3ex\hbox{$>$\kern-.75em\lower1ex\hbox{$\sim$}}\ }
\def\gl{\ \raise.5ex\hbox{$>$}\kern-.8em\lower.5ex\hbox{$<$}\ }
\def\roughly#1{\raise.3ex\hbox{$#1$\kern-.75em\lower1ex\hbox{$\sim$}}}

\def\nn{\nonumber}
\def\at{a_i \tau_i}
\def\spa{\phantom{a}}
\def\pref#1{(\ref{#1})}
\def\hf{\frac12}
\def\del{\partial}

\def\cv{{{\cal{V}}}}

\def\mpl{M_{\rm pl}}

\newcommand\vo{\mathcal{V}}
\newcommand{\comments}[1]{}

\newcounter{oldcounter}
\addtocounter{equation}{1}
\title{ Moduli Vacuum Misalignment and Precise Predictions in String Inflation }
\author[a,b,c]{ Michele Cicoli,}
\author[d]{Koushik Dutta,}
\author[e]{Anshuman Maharana,} 
\author[c,f]{Fernando Quevedo}

\affiliation[a]{\small Dipartimento di Fisica ed Astronomia, Universit\`a di Bologna, \\ via Irnerio 46, 40126 Bologna, Italy}
\affiliation[b]{\small INFN, Sezione di Bologna, Italy}
\affiliation[c]{\small Abdus Salam ICTP, Strada Costiera 11, Trieste 34014, Italy}
\affiliation[d]{\small Theory Division, Saha Institute of Nuclear Physics, 1/AF Salt Lake, Kolkata 700064, India }
\affiliation[e]{\small Harish Chandra Research Intitute, Chattnag Road, Jhunsi, Allahabad 211019, India.}
\affiliation[f]{\small DAMTP, University of Cambridge, Wilberforce Road, Cambridge, CB3 0WA, UK}
\emailAdd{mcicoli@ictp.it}
\emailAdd{koushik.dutta@saha.ac.in}
\emailAdd{anshumanmaharana@hri.res.in}
\emailAdd{f.quevedo@damtp.cam.ac.uk}

\abstract{The predictions for all the cosmological observables of any inflationary model depend on the number of e-foldings which is sensitive to the post-inflationary history of the universe. In string models the generic presence of light moduli leads to a late-time period of matter domination which lowers the required number of e-foldings and, in turn, modifies the exact predictions of any inflationary model. In this paper we compute exactly the shift of the number of e-foldings in K\"ahler moduli inflation which is determined by the magnitude of the moduli initial displacement caused by vacuum misalignment and the moduli decay rates. We find that the preferred number of e-foldings gets reduced from $50$ to $45$, causing a modification of the spectral index at the percent level. Our results illustrate the importance of understanding the full post-inflationary evolution of the universe in order to derive precise predictions in string inflation. To perform this task it is crucial to work in a setting where there is good control over moduli stabilisation.}

\begin{document}

\maketitle

\section{Introduction}

Cosmological observations in the next decade are very likely to play an  important r\^ole in shaping our understanding of the universe. The adiabatic scale invariant spectrum has given strong credence to the inflationary paradigm; upcoming observations  are going to allow for precision tests and also provide us with the data necessary  to scan through the inflationary model space. Thus on the theoretical front the time is ripe to develop the tools necessary to compute  inflationary predictions precisely.
 
In the inflationary paradigm, both the spectral tilt $n_s$ and the tensor-to-scalar ratio $r$ are determined by the slow-roll parameters at the time of horizon exit of the CMB modes. Since the slow-roll parameters in turn are sensitive to Planck suppressed operators, this necessitates the embedding of inflationary models in an ultraviolet complete theory. Thus inflationary model building in string theory has received much attention \cite{R1, R2, R3}.

Moduli fields are a generic feature of string models. They play a central r\^ole in inflationary model building. Moduli vevs appear as parameters in the inflaton potential (which itself can be one  of the moduli), making moduli stabilisation a crucial input for the study of inflationary dynamics. In particular, a precise understanding of the inflaton couplings is necessary to address the $\eta$-problem. 

Another important aspect of the interplay of moduli stabilisation and inflationary dynamics is ``vacuum misalignment" of the moduli fields \cite{cmp, ccmp, cmmp}. During inflation, the inflaton is displaced from its global minimum -- this affects the potential experienced by the moduli fields. Thus the minimum of the potential for the moduli during inflation differs from the minimum during the post-inflationary epoch. This implies that at the end of inflation light moduli (moduli whose post-inflationary mass is less than the Hubble scale during inflation) are typically displaced from their post-inflationary minimum. When the Hubble scale becomes smaller than the post-inflationary mass of the moduli, these fields start to oscillate around their post-inflationary minima with an initial amplitude given by the difference between the inflationary and the post-inflationary minima of the moduli. This leads to an epoch in the post-inflationary history in which the energy density of the universe is dominated by coherent oscillations of the moduli fields. The moduli  eventually decay and the universe reheats. This reheat temperature scales as a positive power of the mass of the displaced modulus. Thus for moduli masses below a certain value the reheat temperature can fall below what is necessary to account for the successes of big bang nucleosynthesis. Thus considerations based on nucleosynthesis lead to a lower bound on moduli masses -- the ``cosmological moduli problem" bound \cite{cmp, ccmp, cmmp}. 

The epoch of modulus domination has another important implication: its effect on the number of e-foldings of the universe $N_e$ between horizon exit of the modes relevant for CMB observations and the end of inflation. Recall that $N_e$ plays a central r\^ole in making inflationary predictions. Given an inflationary potential, the slow-roll parameters at the time of horizon exit can be expressed in terms of $N_e$ by studying the evolution of the scale factor from the point of horizon exit to the end of the inflationary epoch. Since the slow-roll parameters determine $n_s$ and $r$, this can be used to express $n_s$ and $r$ in terms of $N_e$. On the other hand, demanding that the energy density of the universe at the time of horizon exit precisely evolves to the one observed today constrains a particular linear combination of the number of e-foldings in the various epochs in the entire history of the universe (for a detailed discussion see for e.g. \cite{planck14}). 

If one uses a theoretical prior on the history of the universe in the post-inflationary epoch, this constraint turns into an equation for $N_e$ and determines the preferred range of $N_e$ for the cosmology under consideration.\footnote{For the standard cosmological timeline consisting of inflation, reheating, radiation domination, matter domination and finally the epoch of domination by dark energy, this constraint yields $N_e \simeq 55 \pm 5$ \cite{planck14}. For a discussion of the various effects that can cause $N_e$ to deviate from this  ``standard range" see  \cite{Liddle}.} This knowledge of $N_e$ can be used in the expressions for $n_s$ and $r$ in terms of $N_e$ to make inflationary predictions. We emphasise that the above method to determine $N_e$ is sensitive to the post-inflationary history of the universe. In particular, if the post-inflationary history has an epoch in which the energy density is dominated by cold moduli particles, as described in the previous paragraph, this affects the value of $N_e$. Thus the computation of the preferred range of $N_e$ requires a good understanding of the dynamics of the epoch of modulus domination. More specifically, as we will discuss in Sec. \ref{efoldmass}, one needs to have a knowledge of the number of e-foldings $N_{\rm mod}$ for which the epoch of modulus domination lasts.

As described earlier, the epoch of modulus domination arises as a result of ``vacuum misalignment". Thus the magnitude of the ``initial field displacement" caused by  the misalignment is a key input for determining the associated dynamics. Very generic arguments based on effective field theory estimates give the value of this ``initial displacement" to be of order $M_{\rm pl}$ \cite{mismatch, mismatchb, DineR, DineRT, quan}. The other inputs that are necessary to determine $N_{\rm mod}$ are the widths of the moduli fields, as these determine the duration of the epoch. The computation of $N_{\rm mod}$ and its effect on $N_e$ in models where there is a single modulus with post-inflationary mass below the Hubble scale during inflation was carried out in \cite{dm}. The effect of this on the predictions for $n_s$ and $r$ of some prototypical models of inflation was studied in \cite{ddm}. It was found that the change in the preferred range of $N_e$ has a significant effect for inflationary predictions.\footnote{Although it is possible to use  effective field theory estimates for the initial displacement and moduli widths for the purposes of determining the preferred range of $N_e$, it is certainly more appropriate to compute them explicitly -- particularly if one is interested in confronting precision cosmological data.} Motivated by this, in this paper our goal is to carry out an explicit computation for the preferred range of $N_e$ without making use of effective field theory estimates. Both the necessary inputs (the ``initial displacement" and moduli widths) require a good understanding  of the moduli potential and their couplings. The only arena which allows to compute the input information reliably is string compactifications with moduli stabilisation. Thus the requirement of accurate determination of the preferred range of $N_e$ naturally leads us to the study of inflationary models in moduli stabilised string compactifications.

In this paper we will focus on K\"ahler moduli inflation \cite{Kahler} which is a model of inflation in the Large Volume Scenario (LVS) for moduli stabilisation \cite{LVS, LVS2} of  type IIB flux compactifications \cite{GKP}. In this model, the r\^ole of the inflaton is played by a blow-up K\"ahler modulus. On the other hand, the lightest modulus is the volume mode. We explicitly analyse the scalar potential of the theory to compute the vacuum misalignment for the volume modulus in terms of the microscopic parameters of the compactification. We find that the ``initial displacement" of the canonically normalised field is of order $0.1$-$1\, M_{\rm pl}$, in keeping with effective field theory estimates.\footnote{We also study the initial displacement of the other K\"ahler moduli, finding that for blow-up moduli other than the inflaton the initial displacement is of order the string scale $M_s$ (i.e well below $M_{\rm pl}$). This again is in keeping with general expectations since the wave-function of these moduli is localised in the internal dimensions.} 

This, to our knowledge, is the first explicit computation of an initial field displacement - in this sense it is a verification of the effective field theory  arguments  used to obtain the order of magnitude estimate for the initial displacement. Having obtained the ``initial displacement" of the volume modulus we are able to precisely track the post-inflationary history of the universe. This enables us to compute the preferred range of $N_e$ for K\"ahler moduli inflation. We find the central value of the preferred range to be $45$ (below the central value of $50$ in the case of the standard post-inflationary cosmological timeline). Thus the modified post-inflationary history has a significant effect on the inflationary predictions. 

We would like to emphasise that even if, for typical values of the underlying parameters, the volume modulus is heavy enough (has mass of the order $10^8$-$10^9$ GeV) to evade the cosmological moduli problem, it still plays an important r\^ole in determining inflationary predictions. Our results clearly show the importance of taking post-inflationary moduli dynamics into account while making inflationary predictions. Analysis similar to the one carried out in this paper for K\"ahler moduli inflation should be carried out for all string and supergravity models to obtain reliable inflationary predictions.
 
This paper is organised as follows. In Sec. \ref{CM} we review cosmological moduli and the effect they can have on $N_e$, while in Sec. \ref{Seckmi} we review K\"ahler moduli inflation and LVS moduli stabilisation. In Sec. \ref{KMI} we study the precise predictions of K\"ahler moduli inflation by first computing the ``initial displacement" of the volume modulus during inflation, then using it to determine the preferred range for $N_e$ and finally generating the plot which relates $n_s$ to $N_e$. The inflationary prediction for $n_s$ for K\"ahler moduli inflation is obtained from this plot with $N_e$ taken in the preferred range. 

\section{Cosmological Moduli}
\label{CM}

A generic feature of string/supergravity models with moduli masses below the Hubble scale during inflation is a  cosmological timeline characterised by an epoch in which the energy density of the universe is dominated by cold moduli particles. We briefly review this  cosmological timeline and refer the reader to \cite{cmp,ccmp, cmmp, fq, douglas, bobby} for more details. At the end of inflation, light moduli are typically not present in their post-inflationary vacuum --  this occurs since the minimum of the modulus potential during the inflationary epoch differs from the minimum during the post-inflationary epoch. This ``vacuum misalignment" arises because the potential experienced by the moduli fields depends on the value of the inflaton. As the inflaton is displaced from its global minimum during the inflationary epoch, the minimum of the moduli potential is shifted. Generic arguments based on effective field theory principles give this initial field displacement to be of order $M_{\rm pl}$ \cite{mismatch, mismatchb, DineR, DineRT, quan}.

For the purposes of review, let us consider the case when there is a single modulus $\vp$ whose post-inflationary mass $m_{\vp}$ is below the Hubble scale during inflation. At the end of inflation reheating occurs and the universe becomes radiation dominated.\footnote{There is also energy density associated with the modulus displaced from its minimum but at this stage the radiation energy density dominates.} Although the modulus is not at its post-inflationary minimum, the high value of the Hubble constant keeps it pinned at its initial value. With the expansion of the universe, the Hubble constant falls.  The modulus begins to oscillate about its post-inflationary minimum when the Hubble constant falls below the modulus mass. The time average of the  energy density  associated with the oscillating modulus  dilutes as matter, i.e. at a rate significantly slower than that of radiation. If the initial displacement in Planck units is $Y= \vp_{\rm in} \big{/} \mpl > 0.1 \sqrt{ m_{\vp} \big{/} \mpl}$, the  modulus eventually dominates the energy density of the universe. This epoch lasts until the decay of the moduli particles. The universe reheats for a second time with the decay of the modulus. The successes of big bang nucleosynthesis imply  that the second reheat temperature has to be sufficiently high so that the universe was thermal during nucleosynthesis. This requirement yields a lower bound on the mass of order $m_{\vp} \gtrsim 50$ TeV \cite{cmp, ccmp, cmmp, Kawasaki}. This is known as the cosmological moduli problem bound. 

The above cosmological timeline is often called the modular cosmology timeline which is caused by the non-trivial initial field displacement.\footnote{In models with several light moduli there might be multiple epochs of moduli domination. However, when there is a separation of scales between the mass of the lightest modulus and the mass of other moduli, the lightest modulus outlives the others and sets the time scale for the epoch of modulus domination. The dynamics of the system can then be effectively described by a model with a single modulus with the effect of the heavy moduli being incorporated in the reheating epoch after inflation. In models in which there is no distinct lightest modulus, the dynamics is more complicated to analyse. This was discussed briefly in \cite{dm}. In this paper we will confine ourselves to situations in which there is a distinct lightest modulus.} As mentioned in the introduction, a detailed understanding of the associated post-inflationary dynamics of the moduli is necessary in order to obtain the preferred range of $N_e$. We will carry this out explicitly in Sec. \ref{shiftsec} for K\"ahler moduli inflation.

\subsection{Range of e-foldings in Modular Cosmology}
\label{efoldmass}

As discussed in the introduction, the number of e-foldings between horizon exit and the end of inflation plays a central r\^ole in making predictions in inflationary cosmology. Given the inflaton potential, the slow-roll parameters (like $n_s$ and $r$) can be expressed in terms of $N_e$ by tracking the evolution of the scale factor during the inflationary epoch. Thus the knowledge of $N_e$ allows one to make inflationary predictions.

\begin{figure}[!ht]
\centering
\includegraphics[height=60mm,width=90mm]{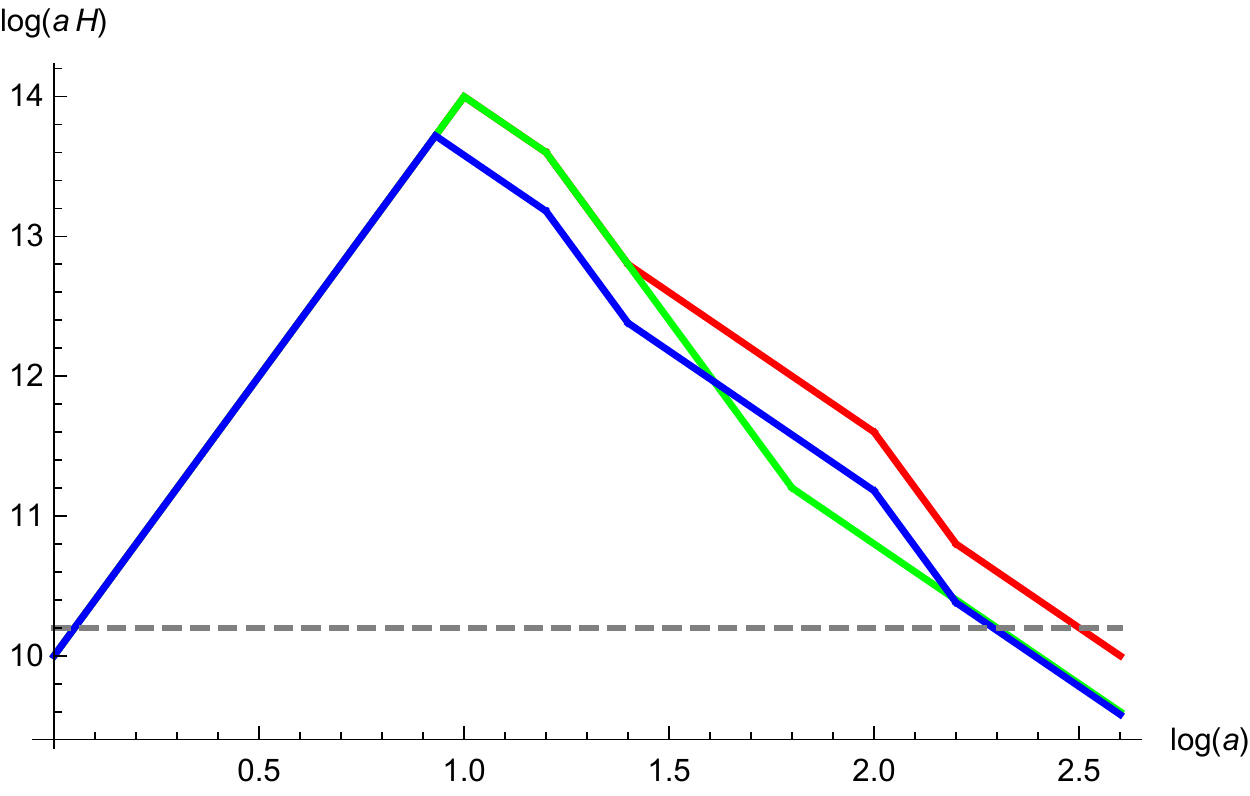}
\caption{Comoving horizon $(a H)^{-1}$ as a function of the scale factor $a$ (the scale is arbitrary). The green line represents a standard cosmological evolution whereas the blue and red lines describe the cosmological evolution of our universe in the presence of light moduli. The red history is inconsistent with present cosmological observations.} 
\label{Fig1}
\end{figure}

The preferred range for $N_e$ is determined by tracking the evolution of the energy density of the universe from horizon exit to the present epoch. The amplitude of the adiabatic perturbations generated by quantum  fluctuations in single field models of inflation is:
\be
A_s = {2 \over 3 \pi^2 r} \left( {\rho_* \over M_{\rm pl}^{4} } \right),
\ee
where $\rho_*$ is the energy density of the universe at horizon exit for CMB modes with wavenumber $k \approx 0.05 \spa {\rm Mpc}^{-1}$. The amplitude $A_s$ remains constant until horizon re-entry and can be related to the CMB temperature fluctuations. Thus the measurement of the strength of temperature fluctuations gives us the value of the energy density of the universe at horizon exit (modulo $r$). We also know the energy density today $\rho_0$ via determination of the Hubble constant. Thus any theoretical proposal for the history of the universe between horizon exit and the present epoch must be such that $\rho_*$ evolves to $\rho_0$. Applying this consistency condition to the standard cosmological timeline (consisting of inflation, reheating, radiation domination, matter domination and the present epoch of acceleration) yields (see for e.g. \cite{planck14}):
\bea
\label{general5}
N_e   + {1 \over 4}( 1 - 3w_{\rm re}) N_{\rm re}  \approx 57  +\frac{1}{4}\ln r    + { 1 \over 4 } \ln \left(\frac{{ \rho_*}} {\rho_{\rm end}} \right),
\eea
where $N_e$ is the number of e-foldings between horizon exit and the end of inflation, $w_{\rm re}$ the effective equation of state parameter during the reheating epoch, $N_{\rm re}$ the number of e-foldings during the reheating epoch which lasts from the end of inflation until the decay of the inflaton, $\rho_*$ the energy density at the time of horizon exit and $\rho_{\rm end}$ the energy density at the end of inflation. Making generic assumptions regarding the reheating epoch, change in the energy density of the universe during inflation and the scale of inflation, ref. \cite{planck14} used \pref{general5} to find the following preferred range for $N_e$ in the standard cosmological timeline:
\bel{raa}
N_e = 55 \pm 5\,.
\ee
As discussed in the introduction, the determination of the preferred range of $N_e$ requires the post-inflationary cosmological history as an input. Thus one expects the preferred range of $N_e$ in modular cosmology to be different from the usual range \pref{raa}. Ref. \cite{dm} applied the above mentioned consistency condition to the modular cosmology timeline described in this section, finding the relation:\footnote{Our notation is slightly different from that of \cite{dm} which used $N_k$ to denote the number of e-foldings between horizon exit and the end of inflation}
\bea
\label{general3}
N_e +{1 \over 4}{N_{\rm mod} }  +  {1 \over 4}( 1 - 3w_{\rm re}) N_{\rm re} \approx 57  +\frac{1}{4}\ln r    + { 1 \over 4 } \ln \left(\frac{{ \rho_*}} {\rho_{\rm end}} \right),
\eea
where $N_{\rm mod}$ is the number of e-foldings that the universe undergoes during the epoch of modulus domination. This corresponds to a second reheating epoch where the equation of state parameter is $w_{\rm mod}=0$.\footnote{We work under the assumption of sudden thermalisation of the modulus decay products. This is a very good approximation since the moduli decay when $H\sim \Gamma_{\rm mod}$. Given that the moduli are only gravitationally coupled while the decay products have gauge interactions with width $\Gamma_{\rm gauge}$, we have $\Gamma_{\rm gauge} > \Gamma_{\rm mod}$. Thus when the modulus decays we have $\Gamma_{\rm gauge}>H$, ensuring a very fast thermalisation process. Note that in the version of eq. \pref{general3} derived in \cite{dm}, a term which captures the effect of this thermalisation epoch was incorporated. As argued above, here we drop this term since its inclusion has a negligible effect in the determination of the preferred range of $N_e$.} The number of e-foldings of modulus domination was found to be:
\bel{Modd}
N_{\rm mod} \approx {4 \over 3} \ln  \bigg( { \sqrt{16 \pi} M_{\rm pl} Y^{2}  \over m_{\varphi} } \bigg)\,,
\ee
where $Y$ is the initial displacement of the modulus from its post-inflationary minimum in Planck units.  Eq. \pref{general3} can be used to obtain the ``preferred range" of $N_e$ for modular cosmology. Making the same generic assumptions as in \cite{planck14}, eq. \pref{general3} gives the preferred range for $N_e$ to be:
\bel{pr}
\bigg(55 - {1 \over 4}{N_{\rm mod}} \bigg) \pm 5\,.
\ee
Note that this can be thought of as a lowering of the central value of the preferred range of $N_e$ by $N_{\rm mod}/4$. This can be clearly seen in Fig. \ref{Fig1} where the comoving horizon is plotted as a function of the scale factor. The green line represents a standard cosmological evolution: inflation, reheating, radiation- and matter-dominance. On the other hand, the blue and red lines represent a cosmological evolution in the presence of moduli: inflation, reheating, radiation-, moduli-, radiation- and matter-dominance. The difference between the blue and the red line is in the duration of
inflation. If inflation in the presence of light moduli lasts as in the standard case (red line), the modes which would be entering the horizon today in a standard cosmology (green line) would still be outside the horizon. In order to make these modes enter the horizon today also in the cosmological evolution
with moduli, inflation has to be shorter (blue line).

As mentioned in the introduction, one can use effective field theory estimates to determine $N_{\rm mod}$ but to compute it explicitly one needs to work in a setting where there is a good understanding of moduli stabilisation. One of the primary goals of this paper is to emphasise the importance of working in a concrete moduli stabilised setting in order to determine the preferred range of $N_e$. We shall take K\"ahler moduli inflation as our model for this purpose. The associated cosmological timeline will be discussed in detail in Sec. \ref{nk}, while here we just note some important features. In the cosmological timeline there are two epochs of modulus domination  -- the first in which the energy density is dominated by inflaton quanta (which are produced during reheating) and the second in which the energy density is dominated by coherent oscillations of the volume modulus. Following \cite{dm}, eq. \pref{general3} is easily generalised to the situation in which there are two epochs of modulus domination:\footnote{Again we work under the good assumption of sudden thermalisation of the moduli decay products.}
\be
N_e +{1 \over 4}{N_{\rm mod1} }  +  {1 \over 4}{N_{\rm mod2} }  \approx 57  +\frac{1}{4}\ln r    + { 1 \over 4 } \ln \left(\frac{{ \rho_*}} {\rho_{\rm end}} \right).
\label{general4}
\ee
Notice that each epoch of modulus domination has a contribution which is equal to one fourth of the number of e-foldings in the epoch. The knowledge of the moduli potential and couplings will provide us with the ingredients (the magnitude of the initial displacement of the volume modulus and the widths of the moduli) necessary to determine the number of e-foldings in the epochs of modulus domination. Another important feature is the contribution from the term involving the tensor-to-scalar ratio $r$ which in K\"ahler moduli inflation is extremely small: $r \sim \mathcal{O}(10^{-10})$. Thus, unlike most models of inflation, the term involving $r$ makes a large contribution in eq. \pref{general4} determining $N_e$.

\section{Review of K\"ahler Moduli Inflation}
\label{Seckmi}

K\"ahler moduli inflation \cite{Kahler} provides a simple realisation of inflation in string compactifications. The model is constructed in the LVS framework \cite{LVS, LVS2} for moduli stabilisation in IIB flux compactifications with a blow-up K\"ahler modulus playing the r\^ole of the inflaton. In IIB flux compactifications the complex structure moduli appear in the Gukov-Vafa-Witten superpotential \cite{gukov}, while the shift symmetry of the K\"ahler moduli implies that they appear in the superpotential only via non-perturbative effects. Thus K\"ahler moduli are ideal candidates for playing the r\^ole of the inflaton -- smallness of the non-perturbative effects can be exploited to obtain the approximately flat potential necessary to satisfy the slow-roll conditions.

\subsection{Large Volume Scenario}
\label{Seclvs}

As discussed in the introduction, the inflationary dynamics in string compactifications is closely tied to moduli stabilisation. We therefore begin by reviewing LVS moduli stabilisation. The Large Volume Scenario provides a very general algorithmic procedure for moduli stabilisation in IIB Calabi-Yau (CY) compactifications with $h^{2,1} > h^{1,1}$ and one of the K\"ahler moduli being a blow-up mode \cite{Cicoli:2008va}. The superpotential takes the form:
\bel{super}
W = \int G_3 \wedge \Omega (z_a)  \phantom{a} +  \sum_{i} A_i\, e^{-a_i T_i}, 
\ee
where $G_3$ is the complex three-form flux, $\Omega(z_a)$ is the CY holomorphic three-form expressed in terms of the complex structure moduli $z_a$ and $T_i$ are the K\"ahler moduli:
\bel{kmouli}
T_{i} = \tau_{i} + i c_i\,,
\ee
where $\tau_i$ are four-cycle volumes and $c_i$ the associated axionic partners. Upon integrating out the complex structure moduli, the first term in \pref{super} gives a constant which we denote by $W_0$:
\bel{dabwlu}
\bigg{\langle} \int G_3 \wedge \Omega \bigg{\rangle} = W_0\,.
\ee
In the LVS framework $W_0$ is an $\mc{O}(1)$ quantity. The tree-level K\"ahler potential is given by (neglecting the terms which depend on the dilaton and complex structure moduli):
\bel{ktree}
K_{\rm tree} = - 2 \ln \bigg(\cv(T_i) \bigg),
\ee
where $\cv$ is the Einstein frame CY volume in string units. The simplest LVS realisations are provided by CY manifolds whose volume takes the Swiss-cheese form:
\bel{sw}
  \cv = \alpha \bigg( \tau_1^{3/2} - \sum_{i=2}^{n} \lambda_i \tau_i^{3/2} \bigg).
\ee
Here $\tau_1$ controls the overall volume. On the other hand, $\tau_2 ,..., \tau_n$ are blow-up modes parameterising the size of holes in the compactification. As in \cite{Kahler} we will confine ourselves to CY manifolds with a Swiss-cheese structure (we refer the reader to \cite{Cicoli:2008va,
Cicoliloops} for LVS realisations in more general CY cases). The leading $\alpha'$ correction to the K\"ahler potential \pref{ktree} is proportional to the CY Euler number $\chi$. After its inclusion $K$ takes the form \cite{BBHL}:
\bel{ka}
K = - 2 \ln \left( \cv + {\hat\xi \over 2} \right),
\ee
where $\hat\xi = { \chi \over 2 (2 \pi)^3 g_s^{3/2}}$. With this, in the regime $\cv \gg 1$ and  $\tau_1 \gg \tau_i \spa ( {\rm{for}} \spa  i >1)$, the superpotential \pref{super} and K\"ahler potential \pref{ka} yield the scalar potential: 
\bel{scalar}
V_{\rm LVS} = \sum_{i=2}^2 { 8 (a_i A_i)^2 \sqrt{\tau_i} \over 3 \cv \lambda_i } e^{-2 a_i \tau_i}
-  \sum_{i=2}^{n} { 4 a_i A_i W_0 \over \cv^2} \tau_i e^{-a_i \tau_i} + {3 \hat\xi  W_0^2 \over 4 \cv^3 }\,.
\ee 
Note that the negative sign of the second term in the above potential arises from minimisation of the axionic fields $c_i$. Minimising the potential \pref{scalar} with respect to $\tau_i$ one finds:
\bel{mincon}
a_i A_i e^{-a_i \tau_i} = { 3 \alpha \lambda_i \over 2 \cv} { ( 1 - \at) \over ( \hf - 2 \at) } \sqrt{\tau_i}\,,
\ee
which motivates the following large volume limit:
\bel{lim}
\cv \to \infty \spa \spa {\rm{with}} \spa \spa a_i \tau_i \approx \ln \cv\,.
\ee
In this limit, the potential for the volume is:
\bel{vpot} 
V_{\rm LVS} = -\frac{3 W_0^2}{2 \mc{V}^3} \left( \sum_{i=2}^n \left[
\frac{\lambda_i \alpha}{a_i^{3/2}} \right] (\ln \mc{V})^{3/2} - \frac{\hat\xi}{2} \right). 
\ee
From the structure of the potential it is easy to see that the potential has an AdS minimum. To obtain an approximately Minkowski minimum (as is necessary to describe the universe in the present epoch), we need to incorporate other ingredients in the effective action. The necessary ingredients can arise from various microscopic phenomena such as anti-D3 branes in warped throats \cite{kklt}, magnetised D7-branes \cite{mag1, mag2}, dilaton-dependent non-perturbative effects \cite{dil} or the effect of D-terms \cite{rum}. The inclusion of such an effect can be captured by the addition in \pref{vpot} of an uplift term of the form:
\bel{upli}
V_{\rm up} = { D  \over \cv^{\gamma} } \spa \spa \spa {\rm{with}} \spa \spa D > 0\,,
\ee
with the value of  $\gamma$ ($1\leq\gamma\leq 3$) depending on the uplift mechanism. The coefficient $D$ has to be tuned so that the potential:
\bel{total}
   V= V_{\rm LVS} + V_{\rm up} = \sum_{i=2}^n { 8 (a_i A_i)^2 \sqrt{\tau_i} \over 3 \cv \lambda_i } e^{-2 a_i \tau_i}
   -  \sum_{i=2}^{n} { 4 a_i A_i W_0 \over \cv^2} \tau_i e^{-a_i \tau_i}    + {3 \hat\xi  W_0^2 \over 4 \cv^3 } + { D  \over \cv^{\gamma} }
\ee
has an approximately Minkowski vacuum. We will discuss the tuning of $D$ in greater detail in Sec. \ref{shiftsec}. We will refer to the global minimum of the potential \pref{total} as the LVS minimum. This will be relevant for describing the universe in the post-inflationary epoch. The masses of the moduli in this vacuum will be important for us later. In the large volume limit, the mass acquired by the small K\"ahler moduli $(\tau_i, i = 2, ..., n)$ is: 
\bel{MT}
m^2_{\tau_i} \simeq{ W_0^2 (\ln \cv)^2 M^2_{\rm pl} \over \cv^2}\,.
\ee
On the other hand, the mass acquired by the overall volume modulus is:
\bel{MV}
m^2_{\cv} \simeq { W_0^2 M^2_{\rm pl} \over {\vo^3 \ln\vo }}\,,
\ee 
making the volume mode the lightest geometric modulus.

\subsection{K\"ahler Moduli Inflation} 

Here we briefly review K\"ahler moduli inflation and refer the reader to \cite{Kahler} for details. The basic idea of K\"ahler moduli inflation  is that one of the ``small moduli" (without loss of generality we will take this to be $\tau_n$) is displaced from its global minimum and plays the r\^ole of the inflaton.  For $e^{a_n \tau_n} \gg \cv^2$ the potential \pref{total} is well approximated by: 
\bel{vap}
V_{\rm inf} = \sum_{i=2}^{n-1} { 8 (a_i A_i)^2 \sqrt{\tau_i} \over 3 \cv \lambda_i } e^{-2 a_i \tau_i}
-  \sum_{i=2}^{n-1} { 4 a_i A_i W_0 \over \cv^2} \tau_i e^{-a_i \tau_i}
+ {3 \hat\xi  W_0^2 \over 4 \cv^3 } + { D  \over \cv^{\gamma} } - { 4 a_n A_n W_0 \over \cv^2} \tau_n e^{-a_n \tau_n}\,. \nn
\ee
The last term is an exponentially flat potential for the inflaton $\tau_n$. The other terms can be thought of as providing a potential for the fields $\cv$ and $\tau_i$ $(i = 2,..., n-1)$. Taking $\tau_i$ to be at their minimum during the inflationary epoch, we obtain:
\bel{vpotwo} 
V_{\rm inf} = -\frac{3 W_0^2}{2 \mc{V}^3} \left( \sum_{i=2}^{n-1} \left[ \frac{\lambda_i   \alpha}{a_i^{3/2}} \right] (\ln \mc{V})^{3/2} - \frac{\hat\xi}{2} \right)  + { D  \over \cv^{\gamma} } - { 4 a_n A_n W_0 \over \cv^2} \tau_n e^{-a_n \tau_n}\,.
\ee
Comparing the above with \pref{vpot} we conclude that the volume direction is not tachyonic and heavy during the inflationary epoch if the ratio:
\bel{ratio}
R \equiv \frac{\lambda_n \,a_n^{-3/2}}{\sum_{i=2}^{n} \lambda_i \,a_i^{-3/2}}  \ll 1\,.
\ee
Notice that for non-perturbative effects generated by gaugino condensation on D7-branes wrapping a four-cycle $a_i = 2 \pi /N_i$, where $N_i$ are the ranks of the gauge groups, while for non-perturbative effects from Euclidean D3-branes $a_i = 2\pi$. Thus arbitrarily small values of $R$ are unnatural. On the other hand, $R \sim 0.1- 0.01$ can be obtained from natural choices of the microscopic parameters. The phenomenological interesting value of the volume modulus for K\"ahler moduli inflation is $\vo \sim 10^5 - 10^6$ \cite{Kahler}.\footnote{We will discuss the derivation of this range in Sec. \ref{Ip}.} It is easy to check that $R \sim 0.1- 0.01$ does not destabilise the potential for values of the volume modulus in this range. Minimising \pref{vpotwo} with respect to $\cv$ we find: 
\bel{ip}
  V_{\rm inf} (\tau_n)  = V_0 - \frac{4
\tau_n W_0 a_n A_n}{\mc{V}_{\rm in}^2}\,e^{-a_n \tau_n} \spa \spa \spa {\rm with} \spa \spa V_{0} = \frac{\beta W_0^2}{\mc{V}_{\rm in}^3}\,, 
\ee
where $\cv_{\rm in}$ is the value of the volume during the inflationary epoch and $\beta$ an $\mc{O}(1)$ constant which we will determine in the next section. The field $\tau_n$ is not canonically normalised. Eqs. \pref{sw} and \pref{ka} imply that in the large volume limit the canonically normalised field $\sigma$ is:
\bel{can}
\frac{\sigma}{M_{\rm pl}} = \sqrt{\frac{4 \lambda}{3 \mc{V}_{\rm in}}} \,\tau_n^{\frac{3}{4}}. 
\ee
In terms of the canonically normalised field the inflationary potential in Planck units is:
\bel{canpot}
V = V_0 - \frac{4 W_0 a_n A_n}{\mc{V}^2_{\rm in}} \left(\frac{3 \mc{V}_{\rm in} }{4 \lambda} \right)^{2/3}  \sigma^{4/3}
\exp \left[-a_n \left(\frac{3 \mc{V}_{\rm in}}{4 \lambda}\right)^{2/3} \sigma^{4/3}\right].
\ee
Note that the scale of inflation is set by $V_0$. As discussed in \cite{Kahler}, the model is very similar to the textbook example of inflation driven by an exponentially flat potential: $V(\sigma)= C_0(1 - e^{-b \sigma})$ (but with a large coefficient $b$ rather than a large field $\sigma$ during inflation). We will discuss the detailed phenomenology of the model in Sec. \ref{Ip}. Finally, we would like to mention that K\"ahler moduli inflation suffers from an $\eta$-problem arising as a result of potential loop corrections to the K\"ahler potential. These corrections may have power law dependence on the inflaton, and so could dominate over the exponentially flat potential \pref{canpot} and lead to a large mass term for the inflaton \cite{fibre, R1}. For the purposes of this paper, we will assume that this problem can be addressed by concrete configurations that suppress these corrections or in general by tuning microscopic parameters of the compactification. We stress that this tuning does not affect our final results on the precise predictions of K\"ahler moduli inflation.

The dynamics of $\tau_n$ (the inflaton) immediately after the end of inflation was analysed in detail in \cite{LK}. It was found that there is non-perturbative production of $\tau_n$ quanta with very high efficiency -- the entire energy density associated with the inflaton is converted into $\tau_n$ quanta within two/three inflaton oscillations around the minimum. The primary decay channel for any of the small moduli is matter on the D7-branes wrapping the associated cycle. Thus unless the inflaton cycle is the same as the cycle supporting Standard Model degrees of freedom, most of the inflaton energy gets dumped into hidden sector fields when the inflaton quanta decay \cite{mazu}. This is not necessarily a problem since, as we have already explained, these degrees of freedom get diluted due the entropy released by the late-time decay of the volume modulus.\footnote{Another potential problem could arise from thermal corrections which could destabilise the volume mode potential, leading to a decompactification limit \cite{Anguelova:2009ht}. However this can be shown to be never the case for K\"ahler moduli inflation \cite{Allahverdi:2016yws}.} Moduli fields usually decay via Planck suppressed interactions, thus their characteristic width is given by:
\bel{tmod11}
\Gamma_{\rm mod} \simeq \frac{m^3_{\rm mod}}{16 \pi M^2_{\rm pl}}\,.
\ee
However the decay of a small cycle K\"ahler modulus to matter fields associated with branes wrapping the cycle provides an exception to this. The wave-functions of the matter fields are localised on the cycle, hence the couplings are suppressed only by the string scale $M_s$. These couplings were determined and the width of $\tau_n$ quanta was computed in \cite{mazu, CQ}:\footnote{Other blow-up modes have similar decay widths.}
\bel{ttau}
\Gamma_{\rm \tau_n} \simeq 0.1 \frac{m^3_{\rm \tau_n}}{M^2_s}\,.
\ee 
On the other hand, the volume modulus decays via Planck suppressed interactions, and so its decay rate is \cite{mazu,dark}:
\bel{tmod1}
\Gamma_{\rm \cv} \simeq \frac{m_\vo^3}{16 \pi M^2_{\rm pl}}\,,
\ee
in keeping with \pref{tmod11}. These widths will play an important r\^ole when we discuss the post-inflationary history of the universe.

\section{Precise Predictions of K\"ahler Moduli Inflation}
\label{KMI}

In this section we derive the precise predictions of K\"ahler moduli inflation by determining the new range for $N_e$ in a non-standard cosmological evolution with a late-time epoch of matter domination due to the presence of light moduli. 

\subsection{Shift of the Volume Mode During Inflation}
\label{shiftsec}

As discussed in the introduction and Sec. \ref{CM}, arguments based on general principles imply that at the end of inflation light moduli find themselves displaced from their post-inflationary minimum as a result of ``vacuum misalignment". This displacement occurs since the potential experienced by the modulus depends on the inflaton; the minimum of the modulus during the inflationary epoch differs from the minimum in the post-inflationary epoch.  The displacement plays a central r\^ole in determining the post-inflationary history of the universe. In K\"ahler moduli inflation, the volume modulus is displaced from its post-inflationary minimum during inflation, hence there is an epoch in the post-inflationary history of the universe where the energy density is dominated by coherent oscillations of the volume modulus.\footnote{We will also compute the displacement of the other K\"ahler moduli. These displacements will be extremely small. This, together with the fact that the other K\"ahler moduli have masses (both in the inflationary and post-inflationary epoch) greater than the Hubble constant during inflation, implies that they relax to their minimum along with the inflaton and hence do not affect the post-inflationary cosmology.} In this section, we explicitly compute this field displacement, finding that the result is in keeping with effective field theory expectations \cite{mismatch, mismatchb, DineR, DineRT, quan}. This initial displacement will be used as an input for the  determination of the preferred range for $N_e$. 

We begin our analysis by determining the coefficient $D$ of the uplift term in the potential \pref{total}. Recall that the value of $D$ is set by the requirement that the post-inflationary vacuum is Minkowski. To impose this condition, we can first use \pref{mincon} to eliminate $\tau_i$ from \pref{total} (the minimisation conditions \pref{mincon} are unaffected by the addition of the uplift term since this term depends only on the volume modulus). After doing so, the potential is a function of only the volume and is given by:
\bel{np}
V =-\frac{3 W_0^2}{2 \mc{V}^3} \left( \sum_{i=2}^n \left[ \frac{\lambda_i \alpha}{a_i^{3/2}} \right] (\ln \mc{V})^{3/2} -\frac{\hat\xi}{2} \right)    + { D  \over \cv^{\gamma} }\,.
\ee      
If $\cv = \cv_{*}$ is the post-inflationary minimum we need to impose:
\bel{condi}
V(\cv_*) = { \del V (\cv_*)\over \del \cv} = 0\,.
\ee
It is useful to define: 
\bel{bet}
\phi \equiv \ln \cv \spa \spa \quad  {\rm and} \spa \spa \quad 
P \equiv \alpha \sum_{i=2}^n \lambda_i\,a_i^{-3/2} = \frac{\alpha}{R}\, \lambda_n\,a_n^{-3/2}\,.
\ee
The conditions in \pref{condi} then read:
\bea
\label{mine}
&\phantom{s}&  - \frac{3 W_0^2}{2 } \,e^{- 3 \st } \bigg( P \st^{3/2} - {\hat\xi \over 2} \bigg)  + D \,e^{-2 \st} = 0\,, \\
\label{nine}
&\phantom{s}&    \frac{3 W_0^2}{2 } \,e^{- 3 \st } \bigg( 3P \st^{3/2} -{ 3 \over 2} P \st^{1/2} - { 3 \hat\xi \over 2} \bigg)  - 2 D \,e^{-2 \st} = 0\,,
\eea
where for simplicity we have set $\gamma =2$.\footnote{It can be easily checked that our results  have very mild sensitivity to the value of $\gamma$.} Combining \pref{mine} and \pref{nine} we have:
\bel{cubic}
\st^{3/2}  -{ 3 \over 2}  \st^{1/2} - {\hat\xi  \over 2P}= 0\,,
\ee
which determines $\st$.\footnote{This implies that to leading order in the large volume expansion $\st \simeq \big( {\hat\xi \big{/} 2P} \big)^{2/3}$.} Making use of the above in \pref{mine} we conclude:
\bel{abc}
D = \frac{9 W_0^2}{4 } P e^{- \st }  \st^{1/2}\,.
\ee
In summary, after adjusting the coefficient of the uplift term so that there is a Minkowski minimum in the post-inflationary epoch, the potential \pref{total}
is given by:
\bel{ul}
 V(\phi) =- \frac{3 W_0^2}{4} e^{- 3 \phi } \bigg(2 P \phi^{3/2} - \hat\xi   - 3P \st^{1/2}e^{(\phi - \st)} \bigg),
\ee 
with $\st$ given in \pref{cubic}.

Let us now examine the potential during the inflationary epoch which is given by \pref{vpotwo}. The last term in \pref{vpotwo} makes a negligible contribution to the potential for the volume in comparison to the others (since $e^{a_n \tau_n} \gg \cv$). Hence the minimum of the volume modulus during inflation is determined by the potential (with $R$ as defined in \pref{ratio}):
\bel{infpote}
V_{\rm in} (\phi) = -  \frac{3 W_0^2}{4} e^{- 3 \phi }  \left[ 2 P \left(1 - R \right) \phi^{3/2} - \hat\xi  
- 3P \st^{1/2}  e^{(\phi - \st)} \right]. 
\ee
This gives that $\phi_{\rm in}$ (the minimum during the inflationary epoch) is determined by:
\bel{newmint}
\big{(} 1 - R \big{)} \phi_{\rm in}^{3/2} -{ 1 \over 2} \big{(} 1 -R \big{)} \phi_{\rm in}^{1/2}    -    e^{(\phi_{\rm in} - \st) }  \st^{1/2} 
- { \hat\xi  \over 2 P}  = 0\,.
\ee
Recall that $\st$ (the minimum of the volume in the post-inflationary epoch) is determined by \pref{cubic} and hence is a function of $\hat\xi \big{/} P$. Thus for a given value of the volume, (\ref{newmint}) determines $\phi_{\rm in}$ as a function of $R$. As discussed in Sec. \ref{Seckmi}, the existence of a stable minimum during inflation requires $R \ll 1$. The shift $\delta \phi = \phi_{\rm in} - \st$ can be obtained by working in a perturbative expansion in this parameter.  For this, it is useful to write the potential during the inflationary epoch as:
\bel{split}
V_{\rm in}(\phi) =  V(\phi) + \delta V (\phi)\,,
\ee
with $V(\phi)$ as given in \pref{ul} and $\delta V(\phi) =  \frac{3 W_0^2}{2 } \,e^{- 3 \phi } P R\, \phi^{3/2}$. The shift in the location of the minimum is then given by:
\bel{shift}
\delta \phi = - {\delta V'(\st) \over V''(\st)} = 4 R \,{\st +{\hat\xi  \over 2P}\,\st^{\hf} \over 2 \st - 1 } \simeq 2 R \st\,,
\ee
where we have made use of the large volume limit in the approximation. Recall that for K\"ahler moduli inflation $\cv_{\rm in} \sim 10^5\tiny{-}10^6$ and for typical values of microscopic parameters $R \sim 0.01 - 0.1$. This gives $\delta \phi \sim 0.1-1$. Note that the volume during the inflationary epoch is greater than the volume in the post-inflationary epoch (since $R > 0$) but it is smaller than the local maximum of the potential (since $R\ll 1$) and therefore the field will roll towards the local minimum and not to the decompactification minimum after inflation. We are interested in the displacement of the canonically normalised field which is $\frac{\varphi}{\mpl} = \sqrt{2 \over 3  } \phi$. Thus we conclude: 
\bel{si}
Y = { \delta \varphi \over \mpl}  = \sqrt{\frac 23} \delta \phi \simeq 2 \sqrt{\frac 23} R \st \simeq 0.1-1\,,
\ee
consistent with the effective field theory expectations based mostly on dimensional analysis. Having obtained the shift in the volume modulus, we can use \pref{lim} to obtain the shift in the other K\"ahler moduli finding: 
\bel{Taushift}
a_i \delta \tau_i \approx \delta \phi \simeq 2 R \st\,.
\ee
Recall that the fields $\tau_i$ are not canonically normalised, while the canonically normalised fields are given by \pref{can}. We can easily see that the displacement of the canonically normalised blow-up modes is of order $\delta\sigma\sim \mpl/\sqrt{\cv}\sim M_s$ (i.e. significantly less than $M_{\rm pl}$). Again, this behaviour is expected as the wave-functions of blow-modes are localised in the internal dimensions. The small initial displacement together with the fact that the blow-up modes (during both inflationary and post-inflationary epochs) are much heavier than the Hubble scale, imply that at the end of inflation they relax to their minimum along with the inflaton and do not have an effect on the post-inflationary dynamics.

Next, let us compute $V_0$ to leading order in $R$. $V_0$ is the expectation value of $V_{\rm inf}(\phi)$ during the inflationary epoch. Since both $V(\phi)$ and its first derivative vanish at $\st$, to leading order in the shift this is given by:
\bel{vz}
 V_0 =  V_{\rm inf}(\phi_{\rm in}) \approx \hf V''(\st) (\delta \phi)^2 + \delta V (\st)  \approx   { 3 W_0^2 \over 2}  e^{-3\st} P R \,\st^{3/2}\,.
\ee
This gives the quantity $\beta$ in \pref{ip} to be: 
\bel{be}
\beta = { 3 \over 2}   P R \,\st^{3/2} = \frac 32 P R \, \left(\ln\vo\right)^{3/2} \,.
\ee
Note that for typical values of the microscopic parameters $\beta$ is an $\mc{O}(1)$ parameter. Eqs. \pref{shift} and \pref{vz} provide expressions for the displacement of the volume modulus and the vacuum energy during inflation to leading order in $R$.
 
\subsection{Number of e-foldings}
\label{nk}

We now have the necessary ingredients to compute $N_e$ in K\"ahler moduli inflation. The dynamics right after the end of inflation at the time $t_1$ was analysed in \cite{LK}. It was found that there is a very violent non-perturbative production of $\tau_n$ (inflaton) quanta, and so the inflaton energy density is converted almost completely into inflaton quanta. Thus the energy density associated with $\tau_n$ quanta is approximately equal to the energy density at the time of inflation computed in \pref{vz}:
\bel{tauen}
\rho_{\tau_n}(t_1)  \approx { M^4_{\rm pl} W^2_{0} \beta \over \cv^3 }\,,
\ee
with $\beta$ as given in \pref{be}. At this stage, there is also energy density associated with the volume modulus arising as a result of vacuum misalignment. This energy density is given by:
\bel{even}
\rho_{\cv}(t_1) \approx  \hf m^2_{\cv} \varphi_{\rm{in}}^2 \approx { M^{4}_{\rm pl}  W_0^2 Y^2 \over \vo^3\ln\vo}\,,
\ee
where $Y$ is the initial displacement in Planck units \pref{si}. Thus the ratio of the two energy densities is given by:
\bel{rat}
\frac{\rho_{\cv}(t_1)}{ \rho_{\tau_n}(t_1)} \approx { Y^2 \over  \beta \ln \cv } \equiv \theta^2\,.
\ee
Since $Y^2 \ll 1$, we have $\rho_{\cv}(t_1)\ll \rho_{\tau_n}(t_1)$ or $\theta^2\ll 1$. The Hubble constant at the end of inflation can be obtained from \pref{tauen}:
\bel{hub}
H(t_1) \approx { M_{\rm pl} W_{0} \beta^{1/2} \over \cv^{3/2} }\,.
\ee
Given that this is of order $m_\vo$, the volume modulus executes coherent oscillations immediately after the end of inflation, and the associated energy density dilutes as matter. Note that since both the energy densities (associated with $\tau_n$ and $\cv$) dilute as matter, the universe has a matter dominated epoch (we will refer to this as the first matter dominated epoch). This also implies that during this period the ratio of the two energy densities remains constant. This epoch lasts until the decay of the $\tau_n$ quanta (that have a shorter lifetime than the volume) which takes place at the time $t_2$. Let us obtain the number of e-foldings of the universe during this epoch. Using the width of $\tau_n$ given in \pref{ttau}, this becomes:
\bel{n1}
N_{\rm mod1}  = \ln\left(\frac{a(t_2)}{a(t_1)}\right) = \frac 13 \ln\left(\frac{\rho_{\tau_n}(t_1)}{\rho_{\tau_n}(t_2)}\right)
\simeq { 2 \over 3} \ln \left(\frac{H (t_1)}{\Gamma_{\tau_n}}\right) 
\simeq {2 \over 3} \ln \left( {10 \beta^{1/2} \cv^{1/2} \over W_0^2 (\ln \cv)^3} \right).
\ee
With the decay of the $\tau_n$ modulus, the associated energy is converted to radiation. However the energy associated with the coherent oscillations of the volume modulus continues to evolve like matter. Note that since the ratio of the energy densities associated with the $\tau_n$ quanta and the volume modulus remains a constant during the first epoch of matter domination, the ratio of the radiation energy density to the energy density associated with coherent oscillations of the volume modulus at $t_2$ is the same as its value at $t_1$ given in \pref{rat}. At this stage, the universe enters an epoch of radiation domination (since $\theta^2 \ll 1$). However, as the universe evolves, the energy density associated with radiation dilutes much faster than the energy density associated with the coherent oscillations of the volume modulus (as the later dilutes like matter), and so the universe eventually enters a second epoch of matter domination which lasts until the decay of the volume modulus. Similar to the estimate for $N_{\rm mod1}$, the number of e-foldings during the second epoch of matter domination is approximately equal to:
\bel{n2}
N_{\rm mod2} \simeq \frac 23 \ln \left(\frac{H (t_{\rm eq})}{\Gamma_\vo}\right),
\ee
where $t_{\rm eq}$ is the time at which equality of radiation and matter energy density (associated with the volume modulus) takes place, while $\Gamma_\vo$ is the lifetime of the volume modulus given in \pref{tmod1}. To determine the Hubble constant at $t_{\rm{eq}}$, first note that \pref{n1} can be used to determine the Hubble constant at $t_2$ in terms of $N_{\rm mod1}$ as:
\bel{hub2}
 H(t_2) = H(t_1) \left( \frac{a(t_1)}{a(t_2)} \right)^{3/2}  = H(t_1)\,e^{-\frac 32 N_{\rm mod1}}
\simeq { H(t_1) W_0^2 (\ln \vo)^3 \over 10 \beta^{1/2} \vo^{1/2}}\,.
\ee
In the subsequent evolution, matter-radiation equality is determined by the condition:
\bel{con2}
\rho_{\rm rad} (t_2) \bigg{(} {a(t_2) \over a(t_{\rm eq})} \bigg{)}^{4} = \rho_\vo (t_2) \bigg{(} {a(t_2) \over a(t_{\rm eq})} \bigg{)}^3\,.
\ee
Since $\rho_\vo (t_2) \big{/} \rho_{\rm rad} (t_2) = \theta^{2}$, this yields $a(t_2)/a(t_{\rm eq}) = \theta^2$. Thus the energy density at the time of equality is $\rho (t_{\rm eq}) \simeq \rho_{\rm rad}(t_2) \,\theta^{8}$ which implies $H (t_{\rm eq}) \simeq H (t_2) \,\theta^4$. Combining this result with \pref{hub2} we obtain:
\bel{hubeqt1}
H(t_{\rm eq}) = { H(t_1) W_0^2 (\ln \cv)^3 \theta^4 \over 10 \beta^{1/2} \ \vo^{1/2}}\,.
\ee
Finally, combining (\ref{tmod1}), (\ref{n2}) and (\ref{hubeqt1}) we obtain: 
\bel{n22}
N_{\rm mod2} \approx {2 \over 3} \ln \bigg( { 16 \pi  \cv^{5/2} (\ln \cv)^{5/2} Y^4  \over 10 \beta^2 }\bigg) 
\approx {2 \over 3} \ln \bigg( { 16 \pi   \cv^{5/2}  Y^4  \over 10 P^2 R^2 (\ln \cv)^{1/2} }\bigg)\,,
\ee
where we have used the expression for $\beta$ as given in \pref{be}. Eqs. \pref{n1} and \pref{n22} determine $N_{\rm mod1}$ and $N_{\rm mod2}$ in terms of the microscopic parameters of the compactification (with the expression for $Y$ given in \pref{si}). These can then be used in \pref{general4} as inputs to determine the preferred range of $N_e$ which, in turn, determines all the inflationary predictions. In particular, in the next section we will obtain the $n_s$-$N_e$ plot.

Before concluding this section, let us estimate the reheat temperature of the universe after the decay of the volume modulus. For this, we need the Hubble constant at the time of decay which is given by:
\bel{RH1}
H(t_{\rm dec}) = H( t_{\rm eq}) \, e^{-\frac 32 N_{\rm mod2}} \simeq {M_{\rm pl} W_0^3 \over 16 \pi \cv^{9/2} (\ln \cv)^{3/2}} \simeq \Gamma_\vo\,.
\ee
The reheat temperature can now be obtained by using:
\bel{RH2}
3 M^2_{\rm pl} H^2 (t_{\rm dec}) =  \rho(t_{\rm dec}) \simeq {\pi^2 \over 30 } g_* \, T_{\rm rh}^4\,,
\ee
where $g_*$ is the effective number of degrees of freedom that thermalise. For the volume in the range $\vo \sim 10^5$ - $10^6$ and $g_* \approx 100$, this gives $T_{\rm rh} \gtrsim 10^3$ GeV showing that there is no tension with the successes of big bang nucleosynthesis.

\subsection{Inflationary Phenomenology}
\label{Ip}

\begin{figure}[h!] 
\centering
\includegraphics[width=8.2cm]{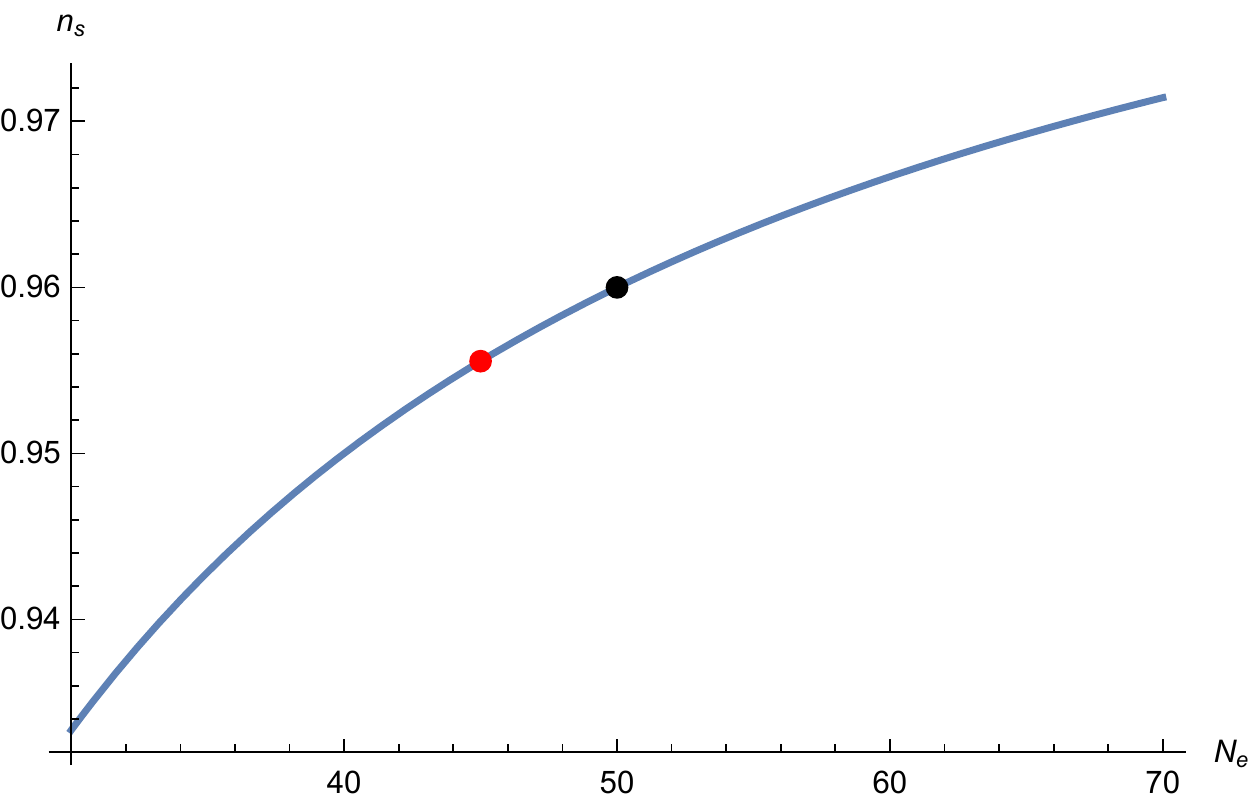} \quad
\caption{Scalar spectral index $n_s$ in terms of the number of e-foldings $N_e$. The black dot shows the value of $N_e$ in a standard cosmological history  while the red dot indicates the value of $N_e$ needed in the presence of a late-time period of modulus domination.}
\label{FignsNe}
\end{figure}

In the inflationary epoch, the field $\tau_n$ rolls down the potential \pref{canpot}. The volume effectively remains at constant value $\mc{V}_{\rm in}$ and the dynamics can be approximated in the framework of single field slow-roll inflation. The slow-roll parameters (which need to be small during the inflationary epoch) are given by \cite{Kahler}:
\be
\label{epsilon}
\epsilon = \frac{M_{\rm pl}^2}{2}\left(\frac{V'}{V} \right)^2 = \frac{32 \mc{V}_{\rm in}^3 }{3 \beta^2 W_0^2 \lambda_n}a_n^2 A_{n}^2\sqrt{\tau_n} \left(1 -  a_n \tau_n\right)^2 e^{-2a_n \tau_n}\,,
\ee
\be
\label{eta}
\eta = M_{\rm pl}^2 \frac{V''}{V} = - \frac{4 \mc{V}_{\rm in}^2 }{3 \beta W_0 \lambda_n \sqrt{\tau_n}}a_n A_{n} \left[\left(1 -  9 a_n \tau_n + 4 a_n^2 \tau_n^2\right) e^{-a_n \tau_n}\right]\,.
\ee
Inflation ends when the slow-roll conditions are violated. This happens when the inflaton reaches the approximate value $a_n \tau_n^{\rm end} \simeq \mc{O} (2 \ln \mc{V}_{\rm in})$. The number of e-foldings as function of the field excursion is given by ($\sigma$ is the canonically normalised inflaton field):
\be
N_e(\sigma) = \int_{\sigma_{\rm end}}^{\sigma}   \frac{1}{\sqrt{2 \epsilon(\sigma)}}\, d\sigma\,.
\ee  
This yields:
\be
 \label{efoldsa}
N_e(\tau_n) = \frac{3 \beta W_0 \lambda_n}{16 \mc{V}_{\rm in}^2 a_n A_n} \int_{\tau^{\rm end}_n}^{\tau_n}\frac{e^{a_n \tau_n}}{\sqrt{\tau_n}(a_n \tau_n-1)}\, d\tau_n\,.
\ee
The integral \pref{efoldsa} can be evaluated exactly in the large volume limit $a_n\tau_n > a_n \tau_n^{\rm end} \simeq \mc{O} (2 \ln \mc{V}_{\rm in}) \gg 1$, finding:
\be
N_e = \frac{3\beta W_0 \lambda_n}{8 \vo_{\rm in}^2 a_n^{3/2} A_n }\left[\frac{e^{a_n \tau_n}}{\sqrt{a_n \tau_n}} +{\rm i} \sqrt{\pi}\, {\rm erf}\left({\rm i}\sqrt{a_n \tau_n}\right)\right]^{\tau_n^{\rm end}}_{\tau_n}\,, 
\label{N-error}
\ee
where ${\rm erf}(x)$ is the error-function. Due to the asymptotic expansion of the error-function:
\be
{\rm i}\sqrt{\pi}\, {\rm erf}\left({\rm i}\sqrt{a_n \tau_n}\right) = -\frac{e^{a_n \tau_n}}{\sqrt{a_n \tau_n}}\left(1+\frac{1}{2 a_n \tau_n}+\cdots \right)\quad\text{for}\,\,a_n \tau_n\gg 1\,,
\ee
the expression (\ref{N-error}) for the number of e-foldings can be approximated as:
\be
N_e = \frac{3\beta W_0 \lambda_n}{16 \vo_{\rm in}^2 a_n^{3/2} A_n }\left[\frac{e^{a_n \tau_n}}{(a_n \tau_n)^{3/2}}\right]^{\tau_n}_{\tau_n^{\rm end}}
\simeq \frac{3\beta W_0 \lambda_n}{16 \vo_{\rm in}^2 a_n^{3/2} A_n } \frac{e^{a_n \tau_n}}{(a_n \tau_n)^{3/2}} \,, 
\label{N-approx}
\ee
where we evaluated the primitive function just at the upper limit of integration due to the presence of an exponential. Substituting (\ref{N-approx}) in (\ref{epsilon}) and (\ref{eta}) for $a_n\tau_n \gg 1$, we obtain:
\be
\epsilon \simeq   \left(\frac{3 \lambda_n}{8 a_n^{3/2}\vo_{\rm in} }\right) \frac{1}{N_e^2 \sqrt{a_n \tau_n}}\qquad\text{and}\qquad 
\eta \simeq - \frac{1}{N_e}\,.
\label{eta2}
\ee
For $a_n\tau_n \gg 1$ we have:
\be
\epsilon \ll   \left(\frac{3 \lambda_n}{8  a_n^{3/2}\vo_{\rm in}}\right) \frac{1}{N_e^2}\simeq \left(\frac{3 \lambda_n}{8 a_n^{3/2}\vo_{\rm in}}\right) \eta^2\,,
\ee
implying that $\epsilon\ll \eta$ for $\eta\ll 1$ and $\vo_{\rm in}\gg 1$. Now matching the COBE normalisation for density fluctuations requires: 
\be
\frac{V^{3/2}}{M_{\rm pl}^3\,V} = 5.2\times10^{-4}\,,
\ee 
which in our set up translates to (including the correct normalisation factor of $V$ as derived in appendix A of \cite{ZB}) :
\be
\left(\frac{g_s\,e^{K_{\rm cs}}}{8\pi}\right)\frac{3 \lambda_n \beta^3 W_0^2}{64 \sqrt{\tau_n} (a_n \tau_n-1)^2}\left(\frac{W_0}{a_n A_n} \right)^2 \frac{e^{2 a_n \tau_n}}{\mc{V}_{\rm in}^6} = 2.7\times 10^{-7}\,,
\label{cobe}
\ee
where $K_{\rm cs}$ is the vev of the tree-level K\"ahler potential for the complex structure moduli. Using (\ref{N-approx}) and working in the limit $a_n\tau_n \gg 1$, (\ref{cobe}) can be rewritten as:
\be
\tau_n \simeq 7.31\cdot 10^{-14} \left(\frac{6\pi\lambda_n}{g_s\,\beta\,e^{K_{\rm cs}}}\right)^2 \left(\frac{\vo_{\rm in}^4}{W_0^4\,a_n^4}\right)\frac{1}{N_e^4} \,.
\label{cobe2}
\ee
After performing a choice of the underlying parameters, (\ref{cobe2}) gives a relation between $\tau_n$ and $N_e$ which guarantees the matching of the COBE normalisation at horizon exit. If we use (\ref{cobe2}) in (\ref{eta2}), we realise that the slow-roll parameter $\epsilon$ does not depend on the number of e-foldings $N_e$ since we obtain:
\be
\epsilon \simeq   3.7 \cdot 10^6 \left(\frac{g_s\,\beta\,e^{K_{\rm cs}}}{16\pi}\right)\left(\frac{W_0^2}{\vo_{\rm in}^3}\right)\,.
\label{epsilon2}
\ee
This implies that the prediction for the tensor-to-scalar ratio $r=16\epsilon$ in this model is independent of the post-inflationary cosmological history. To be more precise, this is true only at leading order since any choice of the underlying parameters has to be such that (\ref{cobe2}) gives $\tau_n\gg 1$. This gives an upper-bound for $r$ that depends on $N_e$:  
\be
r= 16 \epsilon \ll 3.12 \cdot 10^{-3} \left(\frac{6\pi}{g_s\,\beta\,e^{K_{\rm cs}}}\right)^{1/2} \frac{\lambda_n^{3/2}}{W_0\,a_n^3\,N_e^3}\,.
\label{r}
\ee
For $a_n=2\pi$, $g_s \sim 0.1$ and $\mc{O}(1)$ values of $e^{K_{\rm cs}}$, $W_0$, $\lambda_n$ and $\beta$ we obtain:
\be
r\ll 10^{-4}\,N_e^{-3}\,,
\ee
which gives $r\ll 5\cdot 10^{-10}$ for $N_e=60$ and $r\ll 1\cdot 10^{-9}$ for $N_e=40$. We therefore realise that the tensor-to-scalar ratio is undetectable in this inflationary model. On the other hand, the prediction for the spectral index is more interesting since we obtain:
\be
n_s = 1 +2\eta-6\epsilon \simeq 1 -\frac{2}{N_e}\,.
\label{ns}
\ee
This result has two important generic implications: $(i)$ the prediction for the spectral index does not depend on the choice of underlying parameters as long as they satisfy $\tau_n \gg 1$; $(ii)$ models with a smaller number of e-foldings give a smaller value of $n_s$. The $n_s-N_e$ plot is shown in Fig. \ref{FignsNe}. The preferred value for $N_e$ is determined from (\ref{general4}). Typical values of the volume in this model are $\vo_{\rm in}\sim 10^5 -10^6$ which give $r=16\epsilon\sim 10^{-10}-10^{-11}$ from (\ref{epsilon2}) and $N_{\rm mod1}\sim 1$ from (\ref{n1}). Therefore (\ref{general4}) yields (the difference between $\rho_*$ and $\rho_{\rm end}$ gives a negligible contribution):
\be
N_e  \simeq 57  +\frac{1}{4}\ln r -  {1 \over 4}{N_{\rm mod2} }\simeq 50 -  {1 \over 4}{N_{\rm mod2} }\,.
\ee
Note that, as emphasised earlier, the low value of the tensor-to-scalar ratio has a significant effect on determining the preferred range of $N_e$. Let us now come to the effect of the epoch of matter domination due to coherent oscillations of the volume mode. If we neglected this effect, we would need $N_e\simeq 50$ which gives $n_s\simeq 0.96$. However, generic choices of the microscopic parameters in (\ref{n22}) give $N_{\rm mod2}\sim 25$, lowering the number of e-foldings to $N_e\simeq 45$ which, in turn, implies $n_s\simeq 0.955$. This is not a big shift of the spectral index and both results are compatible with Planck 2015 data within $2\,\sigma$ \cite{Planck2015}. However, future cosmological observations \cite{euclid} have the prospect of being able to appreciate this difference. 

We conclude this section by presenting the exact numerical results for a benchmark example. Choosing $W_0=\alpha=\lambda_i=1$, $a_i = 2\pi$ $\forall i=2,...,n=11$ and $g_s = 0.06$ as in ref. \cite{Allahverdi:2016yws}, we find $\vo_{\rm in}\simeq 1.38\cdot 10^5$, $\beta\simeq 3.88$ and a value of $\tau_n$ at the minimum $\langle\tau_n\rangle\simeq 2.26$. Substituting these values in (\ref{cobe2}) and (\ref{epsilon2}) for $e^{K_{\rm cs}}=1$, we find: 
\be
\tau_n \simeq 1.12 \cdot 10^8 \,N_e^{-4} \qquad\text{and}\qquad r=16\epsilon \simeq 1.04\cdot 10^{-10}\,,
\ee
The values of $N_{\rm mod1}$ and $N_{\rm mod2}$ can instead be obtained from (\ref{n1}) and (\ref{n22}), yielding:
\bel{n1n2}
N_{\rm mod1}  \simeq  0.99 \qquad \text{and} \qquad N_{\rm mod2}  \simeq 25.4\,.
\ee
Plugging these results in (\ref{general4}) we obtain: 
\bel{Nkay}
N_e \simeq  44.65  + { 1 \over 4 } \ln \left(\frac{{ \rho_*}} {\rho_{\rm end}} \right)\simeq 45\qquad\Rightarrow\qquad 
\tau_n\simeq 27.3\qquad\text{and}\qquad n_s\simeq 0.955\,.
\ee
In summary, the combined effect of having a low value of $r$ and the epoch of modulus domination is to bring the preferred range of the number of e-foldings to a very low value: $N_e \simeq 45$. Correspondingly, the spectral index becomes $n_s \simeq 0.955$. We would like to emphasise that there is a significant shift in the number of e-foldings $N_e$ even for a heavy volume modulus mass $m_\vo \sim 10^8- 10^9$ GeV. Note that, despite the presence of many parameters ($W_0$, $a_i$, $\lambda_i$, $K_{\rm cs}$), we have been able to extract the relevant information for the region of parameter space that is consistent with observations and obtain precise information on physically measurable quantities such as the spectral index $n_s$.

\section{Conclusions}

In this paper we have studied the inflationary predictions for K\"ahler moduli inflation. To do so, we have determined the preferred range of the number of e-foldings between horizon exit and the end of inflation for the model. This required an analysis of the post-inflationary history of the universe (in particular we determined the number of e-foldings in the epochs of modulus domination). The epoch of modulus domination for the volume modulus results from ``vacuum misalignment". Taking advantage of having knowledge of the moduli stabilising potential in the setup, we have been able to compute explicitly the associated ``initial displacement". Given that the initial displacement is a key input for analysing the post-inflationary history of the universe, being able to compute it explicitly should be considered as an advantage of working in a scenario where there is good control over moduli stabilisation. This we believe is the first explicit computation of ``initial displacement" caused by misalignment. The magnitude of the displacement of the volume modulus agrees with the generic expectations from effective field theory which put it at the order of $M_{\rm pl}$ using mostly dimensional analysis. This has been one of the main assumptions in formulating the cosmological moduli problem, and so our results strengthen the arguments which are used in the discussion of this problem. 

We have found that for K\"ahler moduli inflation the post-inflationary cosmological dynamics of the moduli has a significant effect on the determination of $N_e$ -- its central value is lowered from approximately $50$ to $45$. We would like to emphasise that even if the mass of the lightest modulus in the model is well above the bound set by the cosmological moduli problem ($m_\vo \sim 10^8- 10^9$ GeV), its effect on inflationary predictions via post-inflationary dynamics is still large. The $n_s- N_e$ plot for K\"ahler moduli inflation was given in Fig. \ref{FignsNe}. This clearly exhibits that the shift in 
$N_e$ leads to a shift in $n_s$ at the percent level. Future experiments \cite{euclid} are expected to significantly improve on the sensitivity of $n_s$, making  analysis in the spirit of this paper all the more relevant. This adds to the motivation of studying inflationary model building in moduli stabilised string compactifications. We hope to carry out detailed analysis similar to this paper for various models in future work. 

In this context, a natural model to study is Fibre Inflation \cite{fibre, robust}. In this case, however, there is no epoch of late-time modulus domination since the volume mode decays before the inflaton due to the fact that its mass is larger than the mass of inflaton at the end of inflation. We may also consider extensions of K\"ahler moduli inflation in which the lightest modulus is not the volume mode but a fibre or another non-blow-up modulus field. This field would also be generically misaligned during inflation and behave similarly to the volume modulus in this paper, giving rise to two late-time periods of modulus domination in the post-inflationary history. The combined effect of these two epochs of modulus domination might result in a larger reduction of the required number of e-foldings. It would be interesting to carry out an explicit analysis of the vacuum misalignment of these fields along the lines of this paper. 

Another interesting direction for future work is to carry out a numerical study of the detailed cosmological evolution of the inflaton and the volume mode. This can be used to determine $Y$ (the initial displacement of the volume modulus) without resorting to perturbation theory. It would also be interesting to correlate our findings with other phenomenological implications of modular cosmology \cite{Pheno}. Finally, there have also been studies of non-standard scenarios for primordial fluctuations and non-Gaussianities \cite{ZB,ZB2} in the LVS setting. It would be interesting to study the implications of our analysis for these models.

\section*{Acknowledgements}

We would like to thank Luis Aparicio, Sven Krippendorf and Francesco Muia for useful discussions. KD would like to thank HRI, Allahabad for hospitality. AM would like to thank  ICTP, Trieste for hospitality. KD is partially supported by a Ramanujan Fellowship and a Max Planck Society-DST Visiting Fellowship. AM is partially supported by a Ramanujan Fellowship.

\end{document}